\relax
\documentclass[letterpaper]{article} 
\usepackage{aaai21}  
\usepackage{times}  
\usepackage{helvet} 
\usepackage{courier}  
\usepackage[hyphens]{url}  
\usepackage{graphicx} 
\usepackage{arydshln}
\usepackage{latexsym}
\usepackage{url}
\usepackage{amsmath}
\usepackage{amssymb}
\usepackage{booktabs}
\usepackage{multirow}
\usepackage[switch]{lineno}
\usepackage{natbib}  
\usepackage{caption} 
\title{Deep 3D Mesh Watermarking with Self-Adaptive Robustness}
\author{
    Feng Wang\textsuperscript{\rm 1}, Hang Zhou\textsuperscript{\rm 2},
    Han Fang\textsuperscript{\rm 1}, \\ Xiaojuan Dong\textsuperscript{\rm 1},
    Weiming Zhang\textsuperscript{\rm 1}, Xi Yang\textsuperscript{\rm 1}, Nenghai Yu\textsuperscript{\rm 1}\\
}
\affiliations{
    \textsuperscript{\rm 1}University of Science and Technology of China, \textsuperscript{\rm 2}Simon Fraser University\\
    \textsuperscript{\rm 1}\{nishi@mail., fanghan@mail., xjuadong@mail., zhangwm@, yx9726@mail., ynh@\}ustc.edu.cn\\
    \textsuperscript{\rm 2}zhouhang2991@gmail.com
}

\begin{document}
\maketitle

\begin{abstract}
Robust 3D mesh watermarking is a traditional research topic in computer graphics, which provides an efficient solution to the copyright protection for 3D meshes. Traditionally, researchers need \textit{manually} design watermarking algorithms to achieve sufficient robustness for the actual application scenarios. In this paper, we propose the first deep learning-based 3D mesh watermarking framework, which can solve this problem once for all. In detail, we propose an end-to-end network, consisting of a watermark embedding sub-network, a watermark extracting sub-network and attack layers. We adopt the topology-agnostic graph convolutional network (GCN) as the basic convolution operation for 3D meshes, so our network is not limited by registered meshes (which share a fixed topology). For the specific application scenario, we can integrate the corresponding attack layers to guarantee adaptive robustness against possible attacks. To ensure the visual quality of watermarked 3D meshes, we design a curvature-based loss function to constrain the local geometry smoothness of watermarked meshes. Experimental results show that the proposed method can achieve more universal robustness and faster watermark embedding than baseline methods while guaranteeing comparable visual quality.

\end{abstract}

\section{Introduction}\label{introduction}
With the advent of the new industrial revolution, the 3D industry has become an important industry in society. Therefore, 3D graphics model has became a popular data format in many fields such as arts, games and scientific research. As the dominant 3D shape representation of graphics models, 3D meshes have attracted many researchers in the past few years \cite{fem1997surface,chen2009benchmark}. Since designing and producing 3D meshes is a time-consuming and labor-intensive process, protecting the copyright of 3D meshes has also become a popular task in the 3D mesh industry \cite{copyright2005blind}. Robust 3D mesh watermarking is an efficient solution to this problem. Besides, robust 3D mesh watermarking could serve as an information transmission method. For example, a doctor can insert a patient's personal information into a 3D body skeleton scanned from the diagnostic parts to avoid mismatching the patient and the scanned result \cite{wangComprehensiveSurveyThreeDimensional2008}.

\begin{figure}
\begin{center}
{\centering\includegraphics[width=1\linewidth]{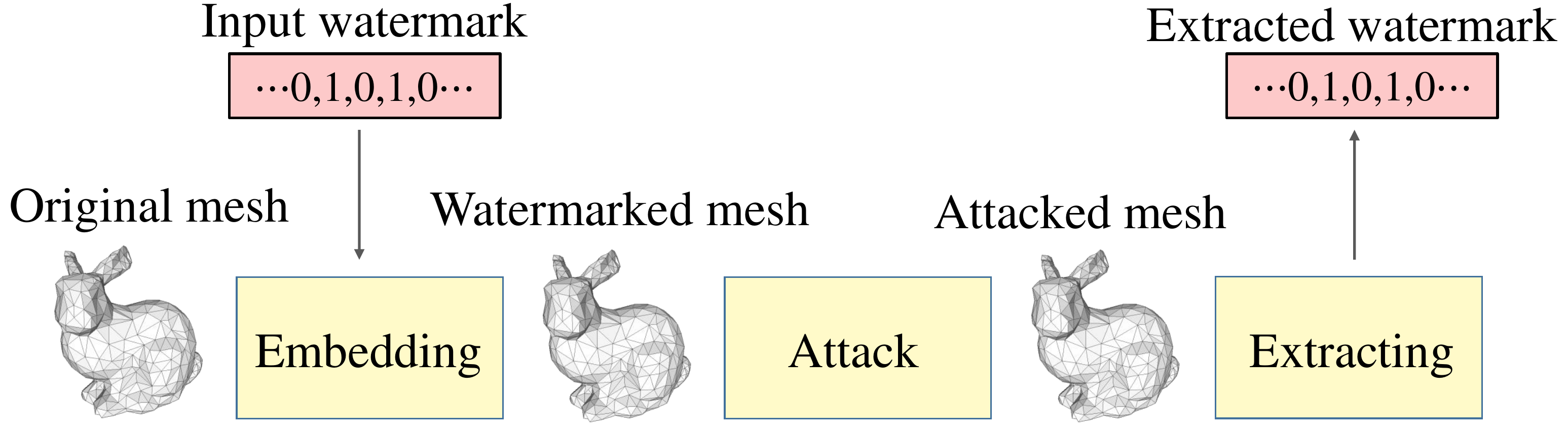}
}\\
\caption{Actual watermarking procedure includes three processes: watermark embedding, attack and watermark extracting. Attack process represents the possible operations on the watermarked mesh.}
\label{fig-framework}
\end{center}
\vspace{-2em}
\end{figure}

Figure \ref{fig-framework} shows the general 3D mesh watermarking model. The watermark represents the message to be embedded. First, the embedding algorithm can embed the watermark into a 3D mesh and generate a watermarked mesh. In actual scenarios, there are many complex geometric and topological operations for 3D data \cite{simply-resilent}, which can cause serious damage to the watermarked mesh. These operations can be regarded as the attack process. For the extracting process, we employ an extracting algorithm to extract the watermark from the attacked mesh. Note that in this paper, we mainly discuss \textit{blind} watermarking techniques, which mean that we can extract the watermark without the reference of the original mesh. And we mainly discuss the methods that embed the watermark by modifying vertices, as these methods are the mainstream methods in this field.

There are the following four requirements for our task. \textit{Robustness}: The watermark should be resilient and not easily removable by possible attacks during the transmission channel. \textit{Imperceptibility}: The visual quality of the watermarked mesh should be guaranteed. \textit{Efficiency}: The time for embedding and extracting the watermark should be as short as possible. \textit{Capacity}: larger capacity indicate that we can embed more message into the mesh. Among all, the most important requirement is robustness, which directly influences the protection ability and transmission accuracy. 

To achieve the above properties, we have to make sufficient efforts to cope with 3D meshes. A 3D mesh can be defined by its vertices and faces, where vertices define the 3D coordinates in the Euclidean space, and faces indicate the topological structure of the mesh. For the data in a 3D mesh file, various attacks may modify it in different ways. These attacks can be divided into three types \cite{chen2009benchmark}: \textit{vertices reordering attack}, \textit{geometry attack} and \textit{connection attack}. Vertices reordering attack can reorder the vertices in the 3D file but doesn't change the 3D coordinates or the topology. Thus it would not change the mesh shape. For the geometry attack, it modifies the vertex coordinates without changing the topological connection. Geometry attacks include similarity transformation, noise addition and smoothing \textit{etc}. Similarity transformation operations consist of three types of transformation: translation, rotation and uniform scaling. Noise addition simulates the artifacts generated during the mesh transmission (e.g. Gaussian noise addition). And smoothing is a common processing operation for 3D meshes, to remove the unevenness of the mesh surface. Contrary to the geometry attack, the connection attack modifies the topological connection between vertices, causing intense damage to the geometric properties of the mesh. Classic connection attacks include cropping, remeshing, simplification and subdivision.

Designing a watermarking algorithm robust against all attacks is impossible. Traditionally, in a specific scenario, to achieve better robustness, we must manually design the specific watermarking algorithm to resist the possible attacks. For example, real-time 3D model rendering needs intense mesh simplification and optimization, which may remove the watermark data \cite{simply-resilent}. As most algorithms are not robust enough against both attacks, we need to develop a specific watermarking algorithm to cope with such a scenario. However, designing a watermarking algorithm for every specific scenario is labor-intensive. What's more, it's difficult to manually design algorithms robust against some attacks, such as the cropping attack. 

To overcome these shortcomings and design a more general watermarking framework for robust 3D mesh watermarking, we propose the first deep learning-based method, which can achieve more universal robustness than traditional methods. In detail, we propose an end-to-end network, consisting of an embedding sub-network, an extracting sub-network and attack layers. Both sub-networks are trained to achieve the watermark embedding and the watermark extracting. And attack layers simulate the actual attacks in the specific scenario. With the differential attack layers, we can jointly train the whole network to find the theoretically optimal solution in the current scenario. For different scenarios, we can adaptively adjust the attack layers to meet the various requirements. Different from 2D image watermarking \cite{zhu2018hidden,tancik2020stegastamp}, 3D mesh watermarking suffers from more threats from 3D space. To guarantee the robustness, we use the topology-agnostic GCN as the basic convolution operation. And consequently, our network can be applied to non-template-based meshes (meshes don't have to share a fixed topology) and the pre-trained model has enough transferability on another remeshed dataset. To better measure the distance between the original mesh and the watermarked mesh, we propose the curvature consistency loss as a constraint for watermarked meshes.

In summary, our main contributions are three-fold:
\begin{itemize}
\item We are the first to introduce a deep learning-based method for robust 3D mesh watermarking task. We hope we can open up new research direction and inspire more works in this field. 

\item We propose a novel deep 3D mesh watermarking network to achieve the adaptive robustness to specific attacks. The curvature consistency loss is proposed to guarantee the visual quality of watermarked meshes. 


\item We quantitatively and qualitatively evaluate the proposed method with two datasets. Experimental results demonstrate that the proposed method can achieve more universal robustness and higher efficiency than baseline methods while guaranteeing comparable visual quality and the same capacity. Besides, our method can be applied to non-template-based meshes, which is very practical in the actual application scenarios.
\end{itemize}

\section{Related Work}

\paragraph{Traditional Robust 3D Mesh Watermarking.} 
Robust 3D mesh watermarking methods can be divided into two categories: spatial domain-based methods \cite{choObliviousWatermarking3D2007,simply-resilent,CG2011,borsOptimized3DWatermarking2013,rolland-neviereTriangleSurfaceMesh2014} and transform domain-based methods \cite{cayreApplicationSpectralDecomposition2003,kanaiDigitalWatermarking3D1998,alfaceBlindWatermarking3D2005,neuralcomputing2017,wang2008hierarchical,ucchedduWaveletbasedBlindWatermarking2004}.

Spatial domain-based methods usually embed the watermark by modifying the spatial parts of a 3D mesh, thus relatively weak to connectivity attack and noise addition attack. And the original structures of 3D meshes can be destroyed by the watermark embedding process, which affects the subsequent mesh synchronization (causality problem). Cho \textit{et al.} \cite{choObliviousWatermarking3D2007} proposed a classic watermarking algorithm based on the distribution of distances between vertices and the mesh gravity center. Before embedding, the vertices are grouped into bins and each bin is assigned with one watermark bit. Based on \cite{choObliviousWatermarking3D2007}, some optimization algorithms are proposed in \cite{borsOptimized3DWatermarking2013,rolland-neviereTriangleSurfaceMesh2014}. The visual quality can be improved but more time is costed during the optimization.



For transform domain-based methods, the common operation is applying the spectral analysis to the original mesh. Then the watermark is embedded by modifying the spectral coefficients of medium frequency parts so that the modification spreads to the spatial components of a mesh. Unfortunately, existing spectral analysis tools have their limitations on the robustness performance against some attacks \cite{wangComprehensiveSurveyThreeDimensional2008}. In \cite{cayreApplicationSpectralDecomposition2003}, Cayre \textit{et al.} first proposed to employ Laplacian matrix as the spectral analysis tool in 3D mesh watermarking task.  Uccheddu \textit{et al.} \cite{ucchedduWaveletbasedBlindWatermarking2004} proposed a wavelet-based watermarking algorithm but the capacity is limited in one bit. And Wang \textit{et al.} \cite{wang2008hierarchical} proposed the hierarchical watermarking algorithm based on wavelet transform. This method can allow for higher capacity, but with weaker robustness.

\paragraph{Deep Learning-based Methods for 3D Mesh Representations.} Different from convolution operation on images, convolution operation on 3D meshes is difficult due to their irregularity and complexity. To lift this limitation, some researches \cite{fengMeshNetMeshNeural2019,meshcnn19,hu2021subdivisionbased,DBLP:conf/nips/MilanoLR0C20,DualConv} have been proposed to effectively learn 3D mesh representation. Yet they can only be applied in discriminative tasks such as classification and semantic segmentation. For generative tasks such as 3D reconstruction, most mesh-based methods use graph convolutional network (GCN) \cite{kipfSemiSupervisedClassificationGraph2017} as the basic convolution operation, where vertices and edges are regarded as nodes and connections in a graph. 

\begin{equation}\label{eq_a_graph_convolution}
f_{i}^{l+1}=\phi(w_{i}f_{i}^{l}+\sum\limits_{j\in\mathcal{N}(i)} w_{j} f_{j}^{l}),
\vspace{0em}
\end{equation}
where $\mathcal{N}(i)$ defines the neighboring vertices of the vertex $i$, $f_{i}^{l}$ is the l-layer feature of the vertex $i$, and $\phi$ is the activation function. Usually, they predict the 3D mesh shape as a deformation from a template \cite{wangPixel2MeshGenerating3D2018,hanockaPoint2MeshSelfPriorDeformable2020}. Besides, a series of efforts \cite{gaoLearningLocalNeighboring2020,gongSpiralNetFastHighly2019,zhouFullyConvolutionalMesh2020} have been proposed to train deep neural auto-encoders to learn latent representations for 3D meshes. These methods usually employ anisotropic filters (each weight $w_j$ variable for every neighboring vertex) to represent the 3D mesh. However, these filters are usually defined based on the fixed vertex order or fixed edge order. 



\section{Proposed Approach}\label{approach}

\paragraph{Topology-Agnostic GCN.}

Due to the possible attacks, watermarked meshes cannot simply be treated as template-based meshes. Even original meshes can also be non-template-based in the actual scenario. To represent these meshes, we employ isotropic filters to compose our convolution operation, with a fixed $w_j$ in Equation \ref{eq_a_graph_convolution} for each neighboring vertex. During training, we find our network converges slowly. We analyze this phenomenon for two reasons: randomly generated watermark bits in each iteration step and different connectivity for each vertex. To speed up training and ensure the convergence, we apply the degree normalization in GCN. We define our GraphConv operation:


\begin{equation}\label{eq_b_graph_convolution}
f_{i}^{l+1}=w_{0}f_{i}^{l}+w_{1} \sum\limits_{j\in\mathcal{N}(i)} \frac{1}{|\mathcal{N}(i)|} f_{j}^{l},
\vspace{0em}
\end{equation}
\begin{figure}[htb]
\begin{center}
{\centering\includegraphics[width=1\linewidth]{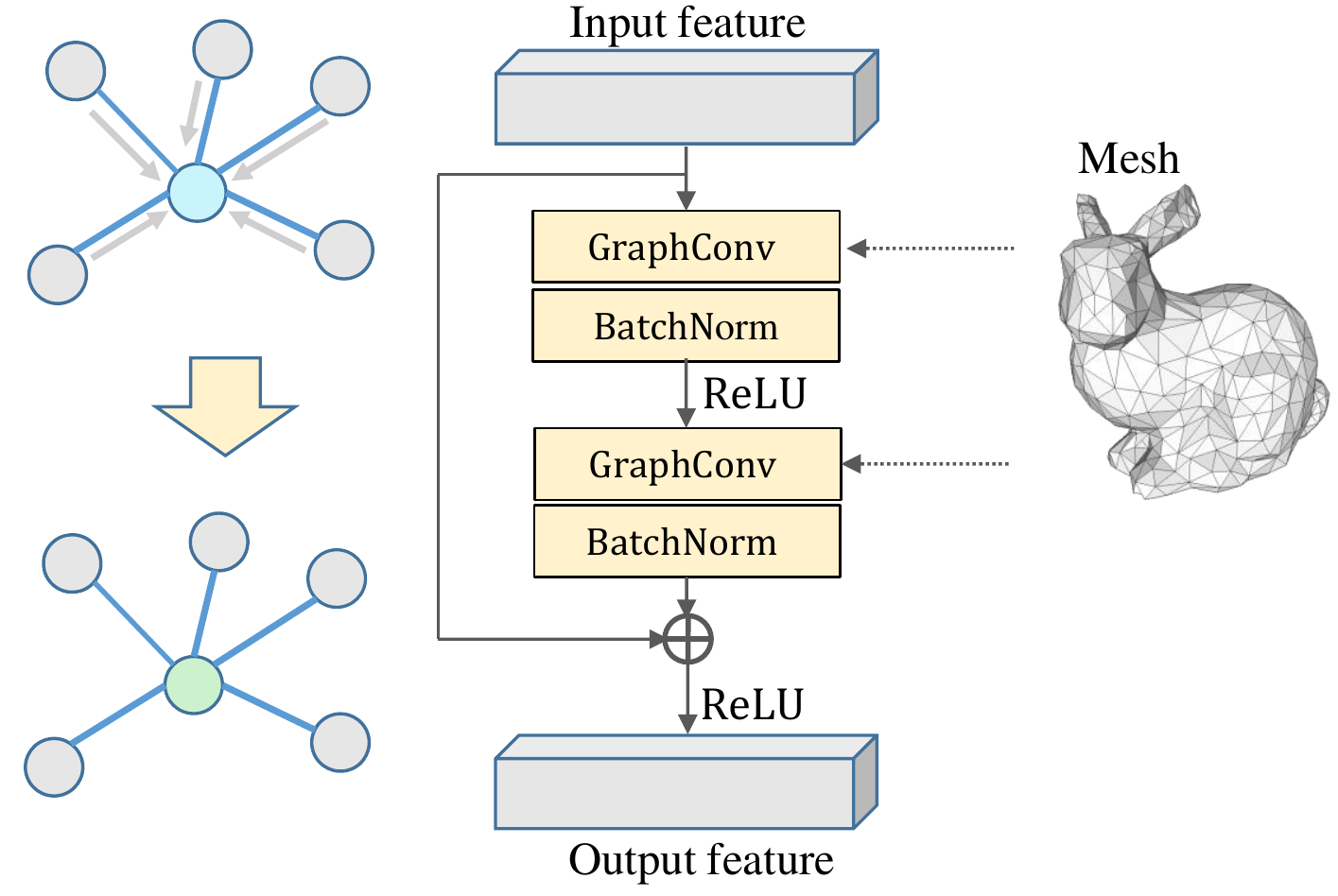}
}\\
\caption{\textbf{Graph residual block}. The dashed line represents that the mesh is utilized for querying the adjacent vertices in GraphConv operation.  }
\label{G-ResBlock}
\end{center}
\vspace{-2em}
\end{figure}
where $|\cdot|$ denotes the cardinal number, indicating the vertex degree. Different from previous GCNs in generative tasks, the topology for each 3D mesh is agnostic. For each mesh with its own topology, topology-agnostic GCN needs to search the neighboring vertices for every vertex. For every mini-batch data, we employ the batch normalization operation to normalize the feature from the output of GraphConv. Shortcut connection could help propagate and aggregate the information from different layers \cite{he2016deep}, so we define the graph residual block consisting of two GraphConv+BN+ReLU blocks with a short connection, as shown in Figure \ref{G-ResBlock}. For the initial block of the embedding sub-network and extracting sub-network, the input feature is the 3D coordinates of vertices and outputs 64-dim feature. For other blocks, the output feature has the same shape as the input feature with 64 dimensions.

\begin{figure*}
{\centering\includegraphics[width=1\linewidth]{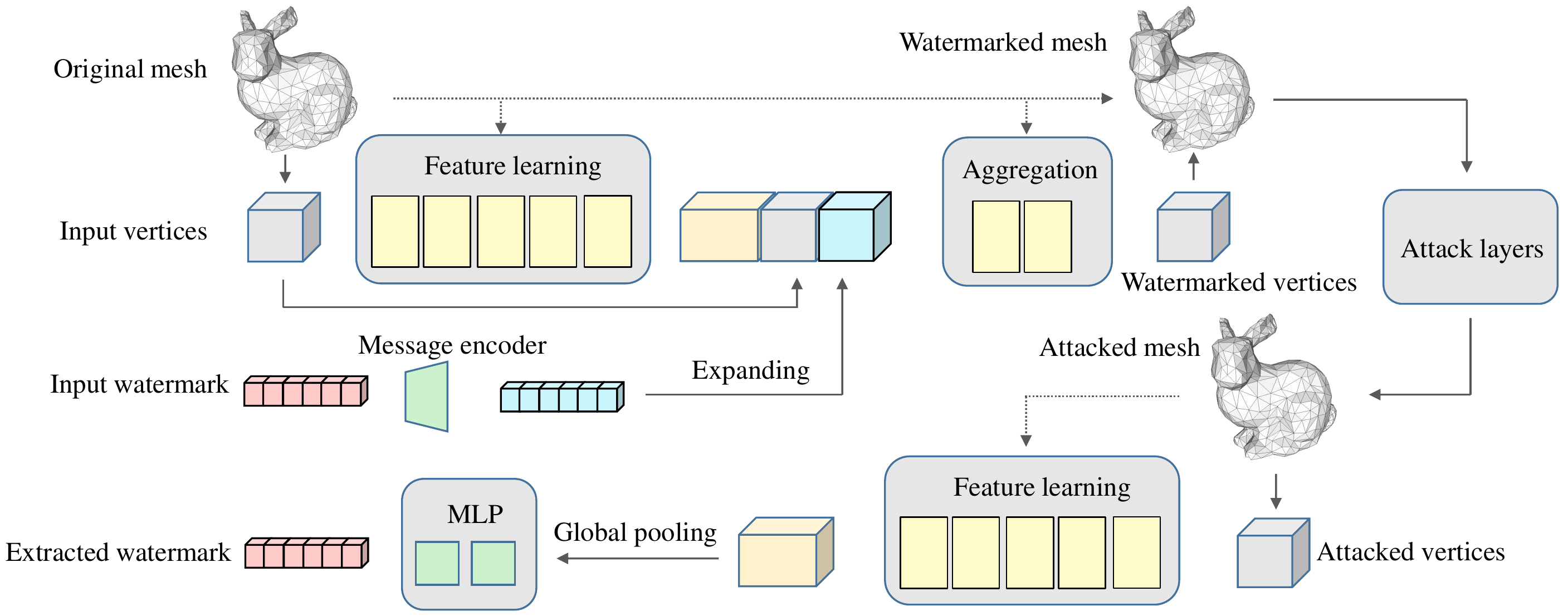}
}\\
\caption{\textbf{Proposed watermark network architecture} for training an end-to-end deep 3D mesh watermarking network. The dashed line represents the reference information that guides the convolution operation and mesh reconstruction.}
\label{fig-E_D}
\vspace{-0em}
\end{figure*}


\vspace{0em}
As shown in Figure \ref{fig-E_D}, our network includes a watermark embedding sub-network, attack layers and a watermark extracting sub-network, corresponding to the watermark embedding process, the attack process and the watermark extracting process respectively. In the network, we define a 3D mesh as $\mathcal{M}=(\mathcal{V}, \mathcal{F})$, where $\mathcal{V}$ denotes vertices and $\mathcal{F}$ denotes faces. And we use $N_{in}$ to denote the number of input vertices. For each vertex $i\in {\mathcal{V}}$, we use $\mathbf{v}_{i}=[x_{i},y_{i},z_{i}]^{\mathrm T}\in \mathbb{R}^3$ to denote the 3D coordinates in the Euclidean space. And we define watermark length as $L$ bits. 
\vspace{0em}

\paragraph{Watermark Embedding Sub-Network.} In this sub-network, we take original mesh $\mathcal{M}_{in}=(\mathcal{V}_{in},\mathcal{F}_{in})$ and watermark $\mathbf{w}_{in}$ as the input. We employ five cascaded graph residual blocks to form the feature learning module $\mathbf{F}$. We first employ this module to learn the feature map $F_{in}$ from input vertices $\mathcal{V}_{in}$. The watermark encoder $\mathbf{E}$ is responsible for encoding the input watermark into a latent code $\mathbf{z}_{w}$ by a fully connected layer. Then the latent code $\mathbf{z}_{w}$ is expanded along the number of vertices to align the vertices. After expanding, the latent code is concatenated with input vertices $\mathcal{V}_{in}$ and the mesh feature $F_{in}$, and then fed into the aggregation module $\mathbf{A}$. In the last block of $\mathbf{A}$, there is a branch that applies an extra GraphConv layer and outputs the 3D coordinates of watermarked vertices $\mathcal{V}_{wm}$. The aggregation module $\mathbf{A}$ includes two graph residual blocks and outputs the 3D coordinates of mesh vertices. According to the original mesh $\mathcal{M}_{in}$ and watermarked vertices $\mathcal{V}_{wm}$, the watermarked 3D mesh $\mathcal{M}_{wm}$ can be constructed. Note that the symmetric function \textit{Expanding} is used to align the vertices and the watermark feature, making the embedding process invariant to the reordering of input vertices, which may be very practical in the actual scenario.



\begin{figure}[htb]
\begin{center}
{\centering\includegraphics[width=1\linewidth]{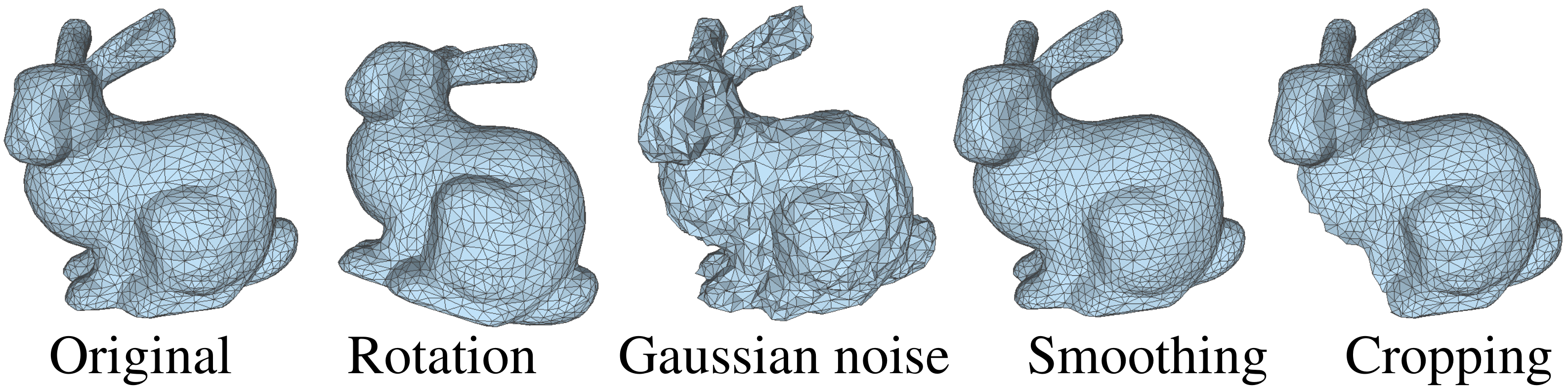}
}\\
\caption{Stanford Bunny model and its attacked meshes.}
\label{noisemesh}
\end{center}
\vspace{-2em}
\end{figure}

\paragraph{Attack Layers.}\label{noise_layers}
To guarantee the adaptive robustness to specific attacks, we train our network with attacked meshes. In this paper, we mainly consider representative attacks (including cropping, Gaussian noise, rotation and smoothing) and integrate them into attack layers. Note that we can integrate different attacks as the attack layers, according to the actual requirements. For the rotation attack, we rotate the 3D mesh in three dimensions with the rotation angle randomly sampled in every dimension. Gaussian noise: We employ a zero-mean Gaussian noise model, sampling the deviation to generate random noise to 3D meshes. Smoothing: We employ a smoothing model \cite{laplaciansmoothing2000geometric} to simulate the possible smoothing operation. Cropping: We simulate this attack by cutting off a part of the mesh. Figure \ref{noisemesh} shows the original and attacked meshes under different attacks. For each mini-batch, we randomly select one of these attacks as the attack layer, applying it to watermarked meshes $\mathcal{M}_{wm}$, then we can obtain the attacked meshes $\mathcal{M}_{att}$. With the differentiable attack layers, we can jointly train our embedding sub-network and extracting sub-network, and update the parameters simultaneously. Details about the attack layers are described in the supplementary material.

\paragraph{Watermark Extracting Sub-Network.}  We design a straightforward structure to extract the watermark. For the attacked vertices $\mathcal{V}_{att}$, we first employ the same feature learning module $\mathbf{F}$ to acquire the feature map $F_{no}$. Followed by the global average pooling layer and a two-layer fully connected layer (MLP), the extracted watermark $\mathbf{w}_{ext}$ is obtained. The symmetric function \textit{Global pooling} aggregates information from all vertices, which can also guarantee the variance under the vertices reordering attack.

\paragraph{Loss Function.}\label{Loss_function}
To train the network, we define some loss functions. Mean square error (MSE) loss is first employed for constraining the watermark and mesh vertices:

\begin{equation}\label{eq_loss_message}
l_{w}(\mathbf{w}_{in},\mathbf{w}_{ext})=\frac{1}{L}\vert\vert\mathbf{w}_{in}-\mathbf{w}_{ext}\vert\vert_2^2,
\vspace{-0em}
\end{equation}

\begin{equation}\label{eq_loss_mesh_vertices}
l_{m}(\mathcal{M}_{in},\mathcal{M}_{wm})
=\frac{1}{N_{in}}\sum_{i\in {\mathcal{V}_{in}}} \vert\vert { {\mathbf{v}_{i} - \mathbf{v}_{i'}}} \vert\vert_2^2,
\vspace{-0em}
\end{equation}
where $i'$ denotes the paired vertex of vertex $i$ in the watermarked mesh $\mathcal{M}_{wm}$.

$l_m$ can constrain the spatial modification on mesh vertices as a whole. Yet the local geometry smoothness is also supposed to be guaranteed, as it greatly affects the visual perception of human eyes. The local curvature can reflect the surface smoothness property \cite{torkhani2012curvature}. For 3D meshes, the local curvature should be defined based on the connection relations. As shown in Figure \ref{curvature}, we use $\theta_{ij}\in [0^{\circ},180^{\circ}]$ to represent the angle between the normalized normal vector $\mathbf{n}_i$ for vertex $i$ and the direction of neighboring vertex $j$. We can find that the vertex's neighboring angles represent the local geometry. For each vertex $i$ in the mesh $\mathcal{M}$, we define the vertex curvature as:

\begin{figure}[htb]
\begin{center}
{\centering\includegraphics[width=1\linewidth]{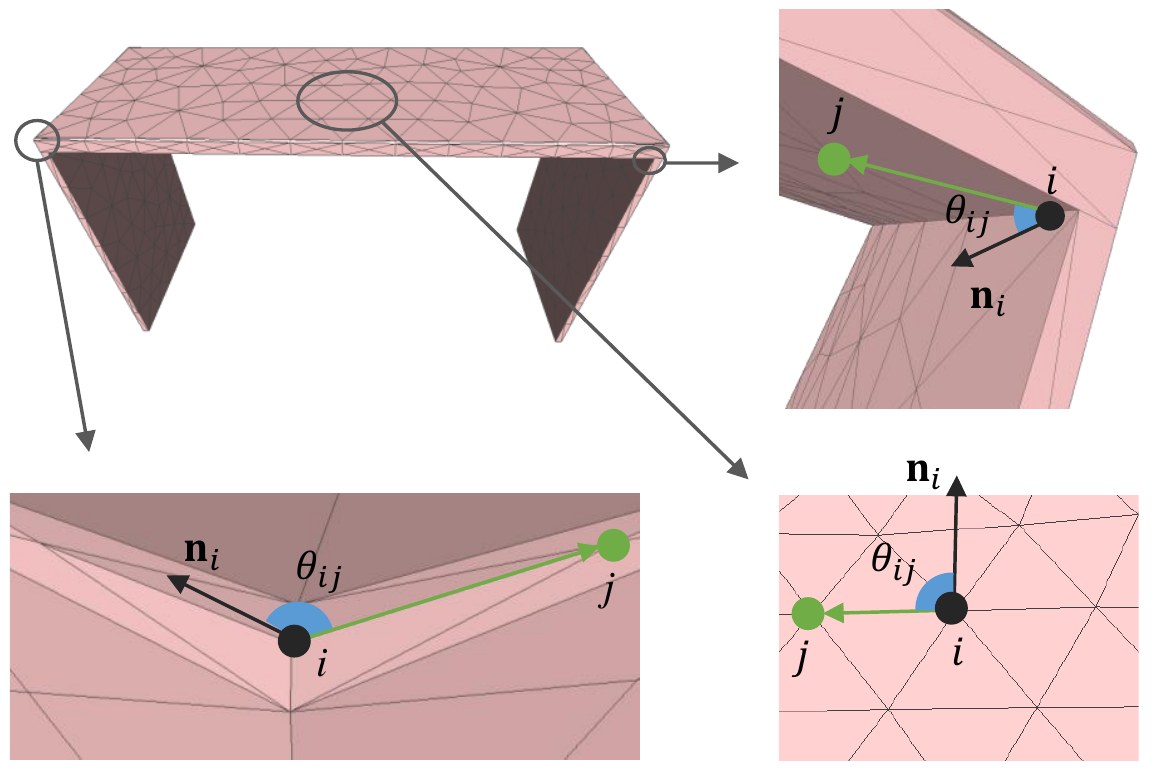}
}\\
\caption{This is a 3D model of a desk. In the bottom right figure, the desktop is flat and the normal vector $\mathbf{n}_i $ is perpendicular to the local area. For each $j\in \mathcal{N}_i$, $\theta_{ij}=90^{\circ} $. The bottom left figure and the top right figure are the convexity and concavity of the the desk respectively, with $\theta_{ij}>90^{\circ}$ and $\theta_{ij}<90^{\circ}$. }
\label{curvature}
\end{center}
\vspace{-2em}
\end{figure}

\begin{equation}\label{eq_vertex_curvature}
cur(i,\mathcal{M}) =\sum_{j\in \mathcal{N}_i} {\rm cos } (\theta_{ij}),
\vspace{-0em}
\end{equation}
where 
\vspace{0em}

\begin{equation}\label{eq_theta}
{\rm cos } (\theta_{ij}) = \frac{(\mathbf{v}_{j} - \mathbf{v}_{i})^{\rm T}\mathbf{n}_i}{\vert\vert \mathbf{v}_j - \mathbf{v}_{i} \vert\vert_2}.
\vspace{-0em}
\end{equation}
To guarantee the local curvature consistency between original 3D mesh $\mathcal{M}_{in}$ and watermarked 3D mesh $\mathcal{M}_{wm}$, we define the curvature consistency loss function:

\begin{equation}\label{eq_loss_curvature}
\begin{array}{ll}
&{l_{cur}(\mathcal{M}_{in},\mathcal{M}_{wm})}\\
\\
&{=\frac{1}{N_{in}}} \sum\limits_{i\in {\mathcal{V}_{in}}}  \vert\vert(cur(i,\mathcal{M}_{in})-cur(i',\mathcal{M}_{wm}))\vert\vert_{2}^{2}.
\vspace{-0em}
\end{array}
\end{equation}
The combined objective is employed in the network: $\mathcal{L} = \lambda_1 l_w +  \lambda_{2} l_{cur} + \lambda_{3} l_{m}$. By default, $\lambda_1=\lambda_{2}=1$, and $\lambda_{3} = 5$.



\section{Experiments}






\paragraph{Implementation Details.}

Our network is implemented by PyTorch and trained on two NVIDIA GeForce RTX 2080Ti GPUs. Adam \cite{experiment2014adam} is applied as the gradient descent algorithm with the learning rate of $0.0001$. We use two scanned datasets: 2D-manifold Hand dataset (triangle meshes with 778 vertices and 1538 faces) \cite{MANO:SIGGRAPHASIA:2017} and 3D-manifold Asiadragon dataset (tet meshes with 959 vertices and 10364 faces) \cite{stanford_graphics}. For Hand dataset, they are divided into 1554 models for train and 50 models for test, and the batch size is 600. For Asiadragon dataset, we use models provided by \cite{zhouFullyConvolutionalMesh2020}, with 7503 models for train and 500 models for test, and the batch size is 400. The network is trained with about one week on both datasets respectively. Before feeding meshes into the network, we normalize vertices to a unit cube. In the experiment, we set the watermark length of all methods as $L=64$.

\begin{figure}[htb]
\begin{center}
{\centering\includegraphics[width=1\linewidth]{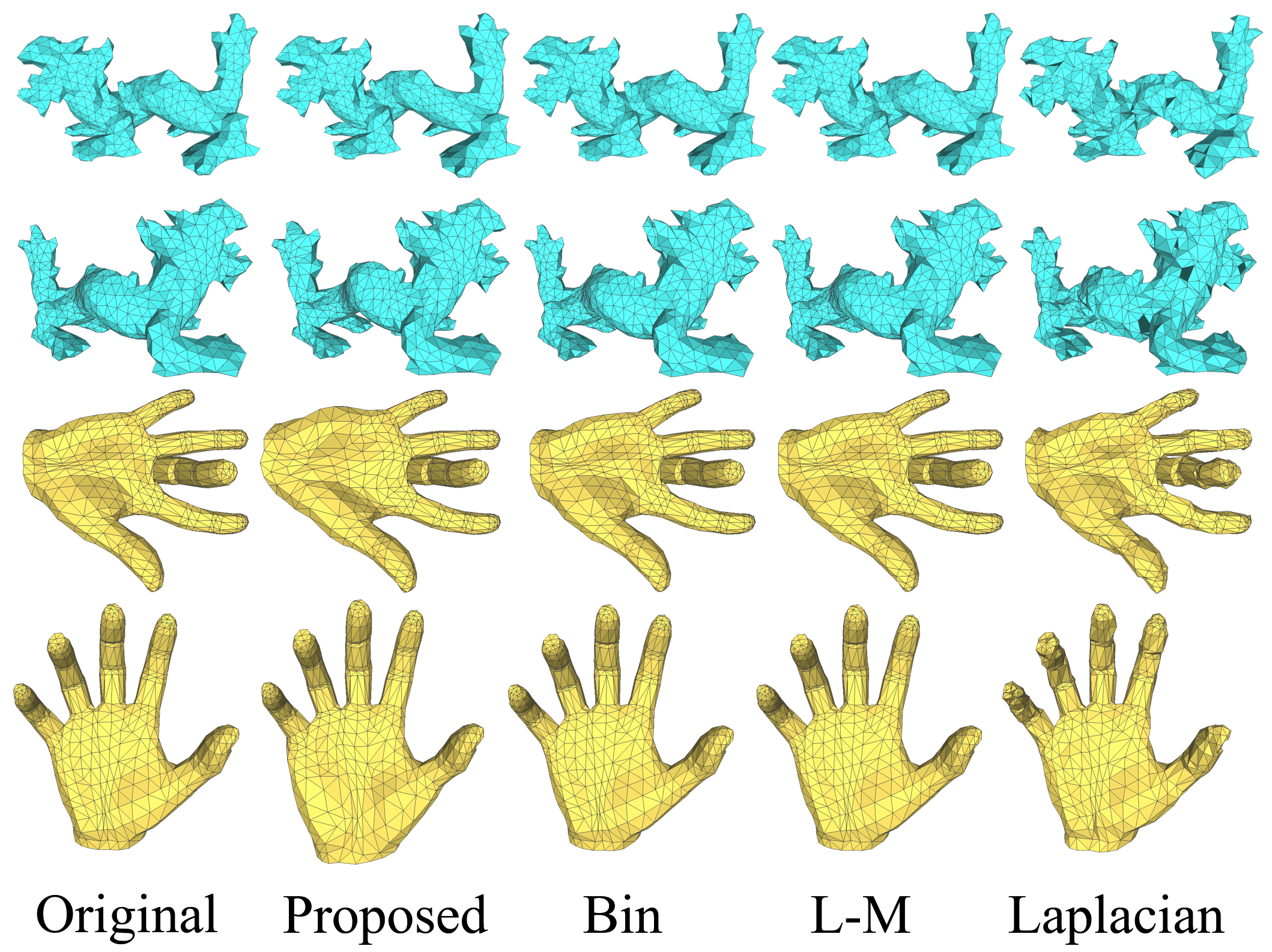}
}\vspace{-1em}
\caption{Qualitative comparison on visual quality with baseline methods on Asiadragon (top two rows) and Hand (bottom two rows) dataset.  }
\label{qualitative-comparison}
\end{center}
\end{figure}

\begin{figure*}[htb]
\begin{center}
{\centering\includegraphics[width=1\linewidth]{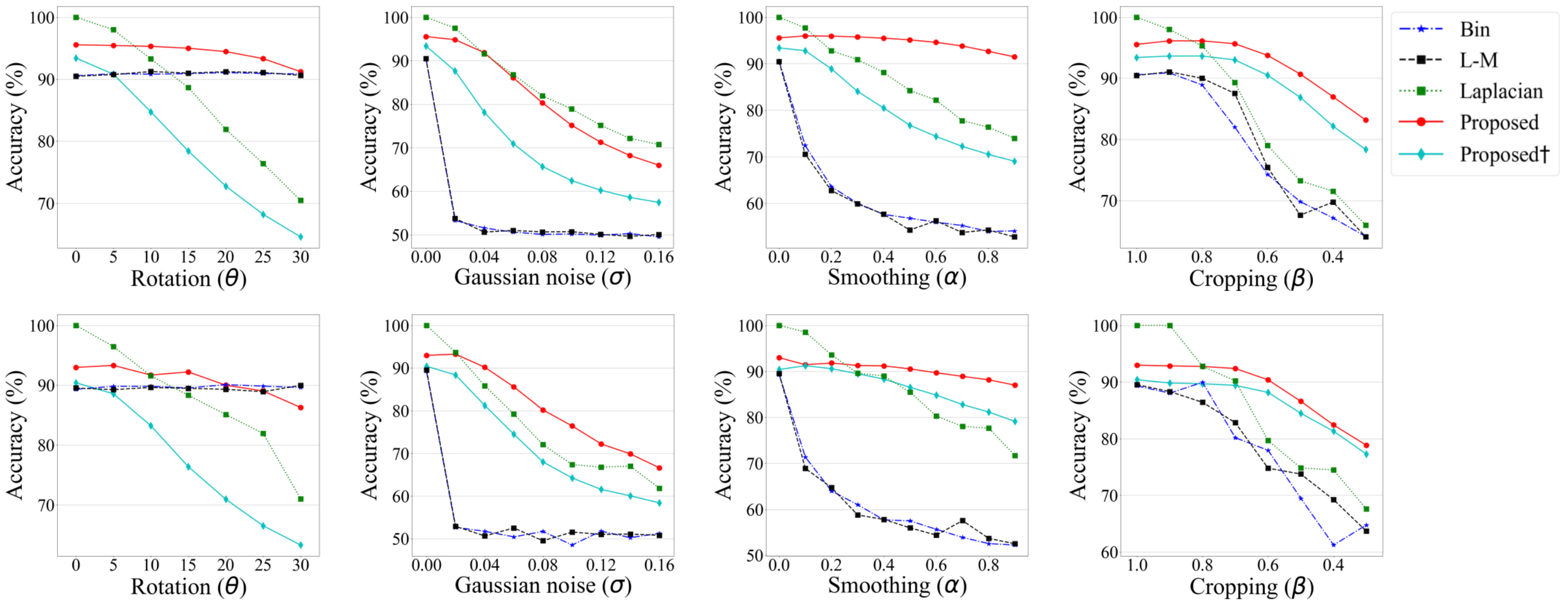}
}\\
\caption{Bit accuracy rate (\%) under different attacks and intensities on Asiadragon (top) and Hand (bottom) dataset. Proposed (red), Proposed$\dag$ (cyan, $\dag$ denotes training without the attack layers), Bin \cite{choObliviousWatermarking3D2007} (blue), L-M \cite{borsOptimized3DWatermarking2013} (black) and Laplacian \cite{cayreApplicationSpectralDecomposition2003} (green) are showed.}
\label{acc}
\end{center}
\vspace{-1em}
\end{figure*}

\paragraph{Evaluation Metrics.}  
We employ Hausdorff distance (\text{HD}), maximum root mean square (\text{MRMS}) and the curvature consistency loss $l_{cur}$  to measure the distances between watermarked meshes and original meshes. To evaluate the robustness, we compare the input watermark bits and extracted watermark bits, and calculate the bit accuracy. Besides, we test algorithms on Intel Xeon Gold 5218 CPU (2.30GHz) and record the mean time consumption for one 3D mesh to compare the efficiency.

\begin{table*}[hbt]
\centering
\scalebox{1}{
\begin{tabular}{cccccccc}
\toprule[1pt]
                            & Method    & HD             & MRMS            & $l_{cur}$        & Accuracy (\%)  & Embedding time (sec) & Extracting time (sec) \\ \midrule
\multirow{4}{*}{Asiadragon} & Bin       & \textbf{0.003} & \textbf{0} & \textbf{0} & 74.47          & 0.406                & \textbf{0.015}        \\
                            & L-M       & \textbf{0.003} & \textbf{0}          & \textbf{0} & 76.62          & 2547                 & \textbf{0.015}        \\
                            & Laplacian & 0.032          & 0.012          & 0.030      & 94.34          & 0.725                & 0.566                 \\
                            & Proposed      & 0.050          & 0.014          & 0.001      & \textbf{95.22} & \textbf{0.032}       & 0.016                 \\ \midrule
                            &           &                &                 &            &                &                      &                       \\ \midrule
\multirow{4}{*}{Hand}       & Bin       & \textbf{0.003} & \textbf{0} & \textbf{0} & 71.84          & 0.327                & 0.013                 \\
                            & L-M       & 0.006          & \textbf{0}         & \textbf{0} & 70.09          & 1040                 & 0.013                 \\
                            & Laplacian & 0.021          & 0.006          & 0.028      & 91.22          & 0.523                & 0.347                 \\
                            & Proposed      & 0.113          & 0.018         & 0.003      & \textbf{92.06} & \textbf{0.022}       & \textbf{0.012}        \\ \bottomrule[1pt]
\end{tabular}

}
\caption{Quantitative comparisons with baseline methods on Hand and Asiadragon dataset. Distance between original meshes and watermarked meshes (second to fourth column), bit accuracy under attacks from the attack layers ($\%$, fifth column) and running time (sixth to seventh column). Running time consists of the watermark embedding time and watermark extracting time for one 3D mesh. For all indicators the lower the better except accuracy.}
\label{quantitative_comparison_dataset}
\end{table*}

\paragraph{Comparisons with Baseline Methods.}

We select three methods as our baseline methods: Bin \cite{choObliviousWatermarking3D2007}, L-M \cite{borsOptimized3DWatermarking2013} and Laplacian \cite{cayreApplicationSpectralDecomposition2003}. In terms of robustness, these methods have the best performances in traditional methods. 

Table \ref{quantitative_comparison_dataset} shows the quantitative comparisons with baseline methods. Tested with the attack layers, the proposed method outperforms Bin by at least $20.75\%$ of accuracy rate and outperforms Bin by at least $18.60\%$ of accuracy rate. In terms of HD and MRMS, the proposed method performs worse than Bin and L-M. That's because they only make minor modifications to the grouped vertices, yet the proposed method need learn the neural representation for 3D meshes, which is currently difficult to achieve the competitive quality as the former. However, as shown in Figure \ref{qualitative-comparison}, we can still keep visually imperceptible. That means HD and MRMS cannot reflect the true visual quality of the mesh. On the contrary, the vertex curvature $cur$ can reflect the local geometry smoothness. With the curvature consistency loss $l_{cur}$ employed during training, the proposed method causes little surface curvature distortion on the watermarked mesh. For Asiadragon dataset, the proposed method get $0.001$ of $l_{cur}$, but Laplacian gets $30\times$ of $l_{cur}$. Besides, we can find that the proposed method can acquire better visual quality than Laplacian. In Figure \ref{qualitative-comparison}, Laplacian causes more distortions on the surface smoothness, making artifacts of watermarked meshes clearly visible. For the efficiency comparison, in Hand dataset, we can find the proposed method only needs 0.022 seconds for the embedding process and 0.012 seconds for the extracting process. Yet other methods cost at least $10\times$ of time for the watermark embedding. 

As shown in Figure \ref{acc}, we test the bit accuracy under each attack with different intensities. L-M and Bin are robust against the rotation attack, but perform badly under other attacks, even with near $50\%$ of accuracy rate under Gaussian noise attack. Laplacian can keep relatively high accuracy under low-intensity attacks, but its accuracy decreases rapidly with the attack intensity increasing. Compared with baseline methods, the proposed method can achieve more universal robustness under all attacks. Although the proposed method cannot guarantee to outperform baseline methods under all conditions, we can still keep the sufficient accuracy under intense attacks, which guarantees the practicality in the actual scenario. For example, we can still obtain the accuracy rate of about $90\%$ on Hand dataset under smoothing attack with $\alpha=0.8$. And under cropping attack with $\beta=0.3$, we have more than $80\%$ accuracy rate on Asiadragon dataset.

\paragraph{The Importance of the Attack Layers.}
As described above, to enhance the robustness against specific attacks, we employ the attack layers during training. To demonstrate the necessity, we also train our network without the attack layers (labelled with $\dag$). As shown in Figure \ref{acc}, we can find that the accuracy decreases a lot under all attacks when training without the attack layers. Under the rotation attack with $\theta=30^{\circ}$, the model training without the attack layers is about $30\%$ of accuracy rate lower than the default model. Under the smoothing attack with $\alpha=0.8$, the accuracy rate is only about $70\%$ in Asiadragon dataset. When training with the attack layers, the accuracy rate can surpass $90\%$.

\vspace{0em}
\paragraph{The Importance of the Curvature Consistency Loss.}

\begin{figure}[htb]
\begin{center}
{\centering\includegraphics[width=1\linewidth]{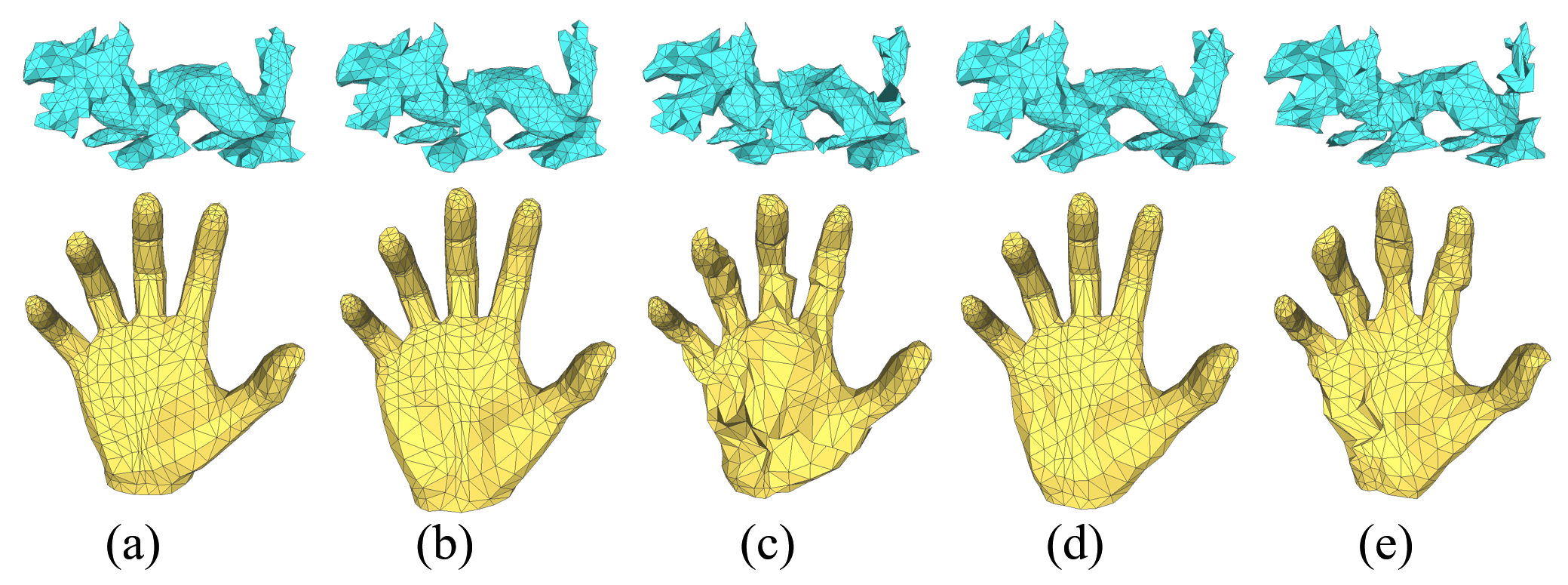}
}\vspace{-1em}
\caption{Visual quality for the importance discussion of curvature consistency loss on Asiadragon (top) and Hand (bottom) dataset. (a) Original; (b) Proposed (w/ $l_{cur}$); (c) Proposed (w/o $l_{cur}$); (d) Proposed$\dag$ (w/ $l_{cur}$); (e) Proposed$\dag$ (w/o $l_{cur}$).}
\vspace{-0em}
\label{cur}
\end{center}
\end{figure}

Besides MSE loss constraining the spatial range of vertices, curvature consistency loss can guarantee the surface smoothness of watermarked meshes. To validate its importance for the visual quality of watermarked meshes, we retrain our models without the curvature consistency loss. As shown in Figure \ref{cur}, we can find that there are many visual artifacts on the watermarked meshed when training without the curvature consistency loss.

\paragraph{Performances on Non-Template-Based Datasets and the Transferability Discussion.}

In the above sections, we mainly discuss the performance of the proposed method on template-based 3D meshes. In the actual scenario, we may need to embed the watermark into non-template-based meshes. To evaluate the proposed method on non-template-based datasets, We independently remesh each shape of Hand and Asiadragon dataset to $1024$ vertices by Trimesh library \cite{trimesh}. Then we retrain our network using remeshed datasets. In the supplementary material, we provide experimental results on these datasets and demonstrate the validity of the proposed method. In addition, the proposed method can also guarantee the transferability of the pre-trained model on the another remeshed dataset. For example, the pre-trained model trained on Hand dataset can still maintain sufficient robustness on remeshed Hand dataset, rather than only $50\%$ of accuracy rate. 

\paragraph{Discussion: How Does Our Network Embed the Watermark into the 3D Mesh?}

\begin{figure}[htb]
{\centering\includegraphics[width=1\linewidth]{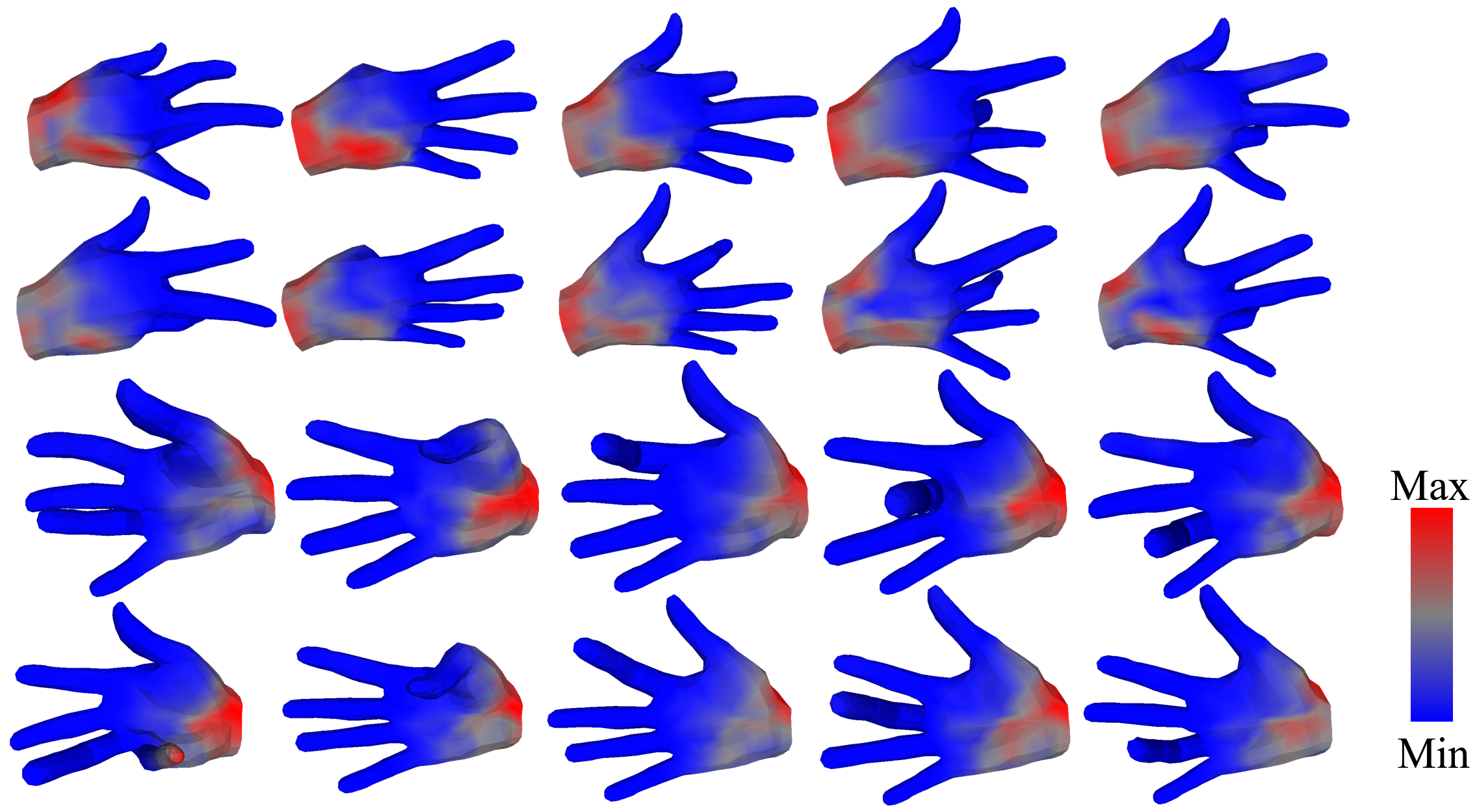}
}
\vspace{0em}
\caption{Colormaps of per vertex Euclidean modication produced on Hand dataset.}
\label{fig-heat}
\end{figure}

Different from traditional methods, we don't know how the network modifies the vertices and embeds the watermark into 3D meshes. Therefore, we explore to analyze the modification based on spatial domain and transform domain. For watermarked vertices and original vertices, we calculate the distances between them in the Euclidean space. Then we color the original 3D mesh based on the $l_2$ distance. As shown in Figure \ref{fig-heat}, we can find that our network prefers to modify vertices on flatting areas, such as the wrist, yet the fingers have fewer modifications. We speculate that there are undulating curvatures in the finger areas, resulting in larger loss from modifications. So the network is trained to prefer to embed the watermark bit in relatively flatting areas.

\vspace{-0em}

\begin{figure}[htb]
{\centering\includegraphics[width=1\linewidth]{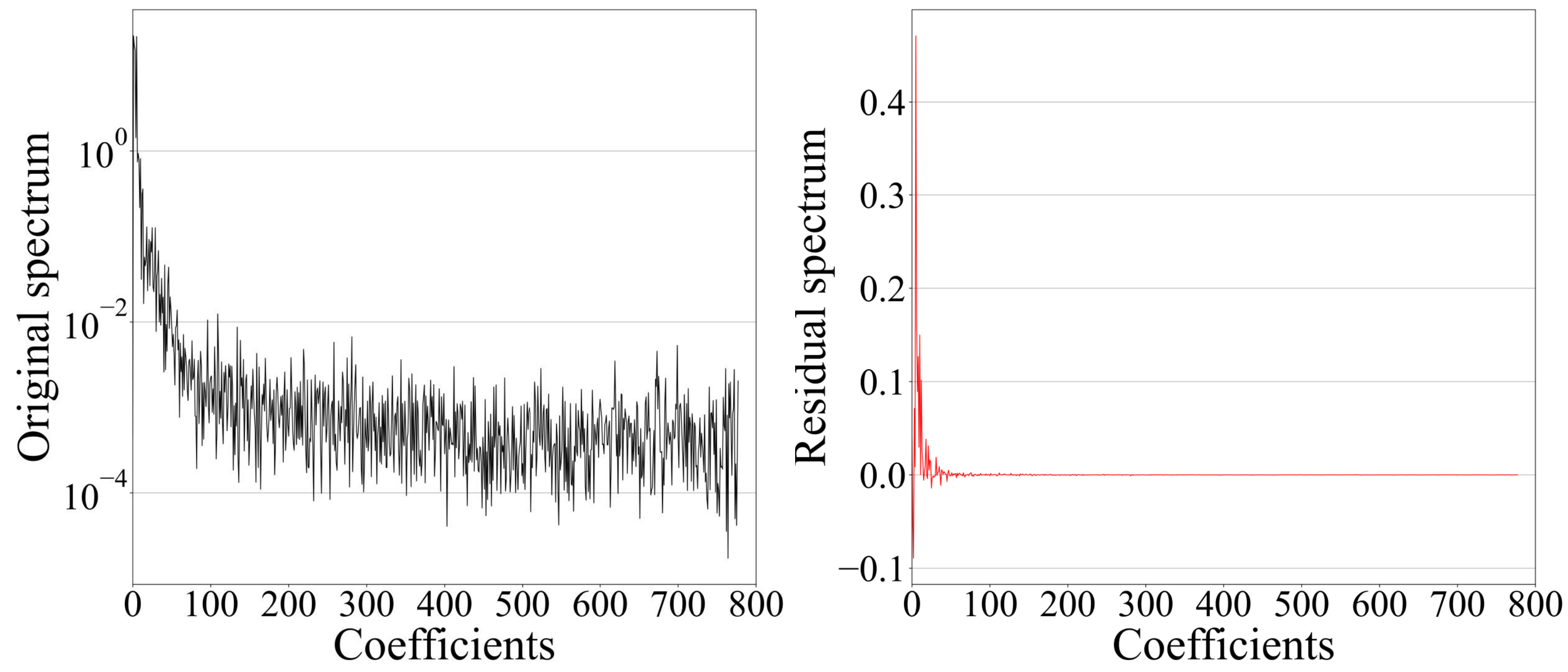}
}
\caption{The original power spectrum (left) and the residual power spectrum (right). Coefficients are ordered with respect to the corresponding eigenvalues of the Laplace-Beltrami operator.}
\label{fig-original}
\vspace{-0em}
\end{figure}

Meanwhile, we perform Laplace-Beltrami operator on Hand dataset and calculate the mean power spectrum of 3D meshes \cite{cayreApplicationSpectralDecomposition2003}. The residual power spectrum between watermarked meshes and original meshes is also calculated. In Figure \ref{fig-original}, low coefficients represent the principal components of the mesh, with higher power spectrum intensity. In the right figure, we find that low coefficients also have more residual power spectrum. That means the network prefers to modify the vertices on the principal components.

\paragraph{Limitations.}



Our experiments are limited in the digital domain and are conducted with several common attacks. To better evaluate the proposed method, we need conduct the experiments in real-world scenarios, such as 3D printing-scanning process \cite{hou2017blind} and 3D-to-2D process \cite{3D2D}. In the future, we will extend our research to these scenarios. 


\section{Conclusion}

In this paper, we propose the first deep learning-based method for the robust 3D mesh watermarking task. We propose a novel end-to-end 3D mesh watermarking network, which can solve this task without manually designing algorithms. Attack layers can improve the robustness against corresponding attacks. In real applications, we can adaptively adjust our attack layers to meet the actual robustness requirement. For visual quality, we design a curvature-based loss function to guarantee the surface smoothness. Extensive experiments demonstrate the effectiveness of our framework and the superior performance of the proposed method.

\bibliography{reference}

\end{document}


\maketitle

\section{Overview}
This document provides technical details and additional experimental results. First, we describe the details of the network architecture and the attack layers. Then we provide additional experimental results on non-template-based datasets. Finally, the importance of the normalization for topology-agnostic GCN is discussed.

\section{Details about the Network Archietcture  \label{network}}

\begin{figure*}[htb]
{\centering\includegraphics[width=1\linewidth]{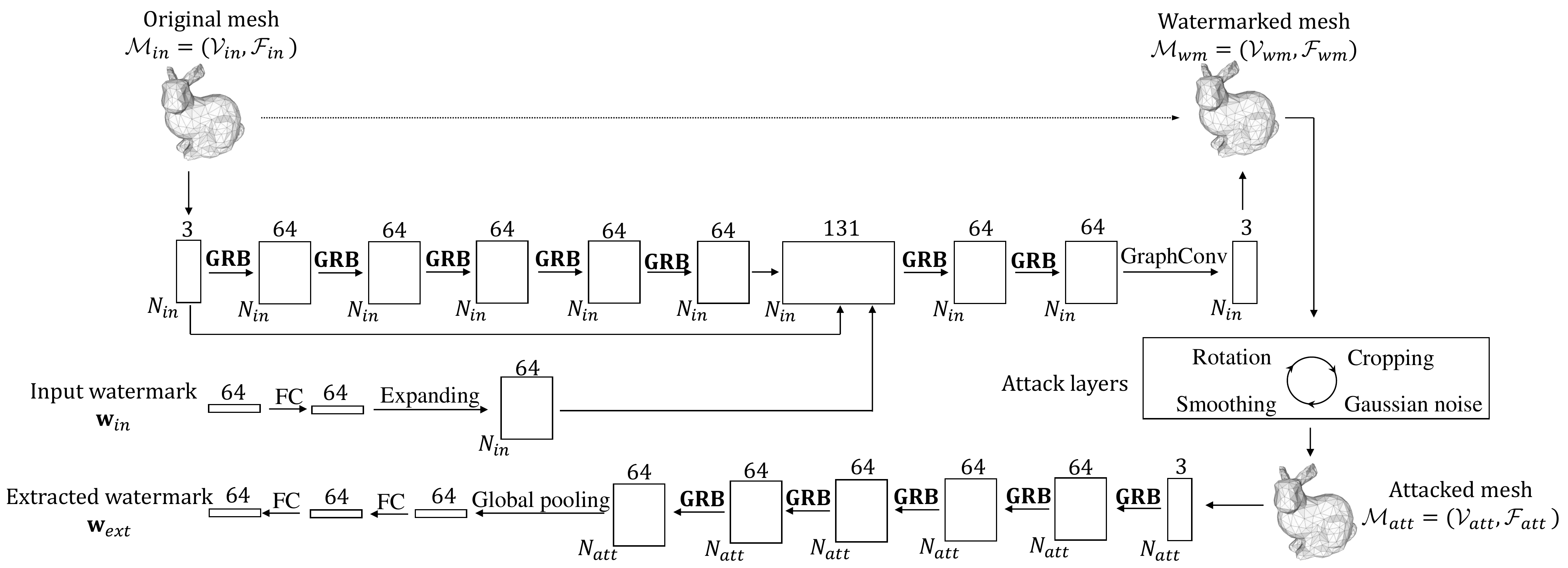}
}\vspace{-0em}
\caption{The detailed illustration of the proposed network. The dashed line represents the reference information for the mesh reconstruction. Note that all graph convolution operations are also guided by the corresponding mesh.}
\label{fig-network}
\vspace{-0em}
\end{figure*}

Figure \ref{fig-network} shows the detailed network architecture in our experiment. \textbf{GRB} represents the graph residual block described in the main text. The input of the whole network is the original mesh $\mathcal{M}_{in}=(\mathcal{V}_{in},\mathcal{F}_{in})$ and input watermark $\mathbf{w}_{in}$ (the watermark length $L=64$). The output of the embedding sub-network is $\mathcal{M}_{wm}=(\mathcal{V}_{wm},\mathcal{F}_{wm})$ and the  output of the extracting sub-network is $\mathbf{w}_{ext}$. We define the input mesh and attacked mesh to have $N_{in}$ and $N_{att}$ vertices respectively. Note that only under cropping attack, $N_{in} > N_{att}$. And under other attacks, $N_{in} = N_{att}$.

\section{Details about the Attack Layers}

In our experiment, we integrate four attacks as the attack layers:

\paragraph{Rotation.} We use $\theta$ to denote the rotation scope and the rotation angle in each dimension is randomly sampled: $\theta_x,\theta_y,\theta_z\sim\textit{U}[-\theta,\theta]$. Then we rotate $\mathcal{V}_{wm}$ with the corresponding angle for every dimension in (x,y,z) space.

\paragraph{Gaussian noise.} We employ $\sigma$ to control the noise intensity. We employ a zero-mean Gaussian noise model, sampling the standard deviation $\sigma_{g} \sim\textit{U}[0,\sigma]$ to generate random noise to 3D meshes. We generate $\textit{noise} \sim\mathcal{N}(0,{\sigma_{g}} ^ {2})$ and attach it on the 3D coordinates of watermarked vertices.

\begin{table*}[htb]
\centering
\scalebox{1}{
\begin{tabular}{cccccc}
\toprule[1pt]
                            & Method            & HD             & MRMS           & $l_{cur}$            & Accuracy (\%)  \\ \midrule
\multirow{4}{*}{Asiadragon} & Model          & 0.059          & 0.014          & \textbf{0.002} & \textbf{93.25} \\
                            & Model$\dag$     & \textbf{0.042} & \textbf{0.010}  & \textbf{0.002} & 84.63          \\
                            & Pre-Model       & 0.069          & 0.016          & 0.006          & 83.63          \\
                            & Pre-Model$\dag$  & 0.058          & 0.012          & 0.004          & 76.51          \\ \midrule
                            &                   &                &                &                &                \\ \midrule
\multirow{4}{*}{Hand}       & Model          & 0.112          & 0.018          & 0.005          & \textbf{91.78} \\
                            & Model$\dag$     & 0.101          & \textbf{0.012} & \textbf{0.003} & 85.09          \\
                            & Pre-Model      & 0.113          & 0.019          & 0.004          & 83.00             \\
                            & Pre-Model$\dag$ & \textbf{0.099} & \textbf{0.012} & \textbf{0.003} & 77.34          \\ \toprule[1pt]
\end{tabular}
}
\caption{Quantitative results on remeshed Hand and remeshed Asiadragon dataset. Distance between original meshes and watermarked meshes (second to fourth column), bit accuracy under attacks from the attack layers ($\%$, last column). $\dag$ denotes training without the attack layers, Model represents the model trained on the remeshed dataset, and Pre-Model represents the model trained on the original dataset.}
\label{quantitative_comparison_remesh}
\vspace{-0em}
\end{table*}

\paragraph{Smoothing.} Laplacian smoothing model is employed to simulate the possible smoothing operation. For the watermarked mesh $\mathcal{M}_{wm}=(\mathcal{V}_{wm},\mathcal{F}_{wm})$, we first calculate the Laplacian matrix $\mathbf{L} \in \mathbb{R}^{N_{in} \times N_{in}}$. For each element $L(i,j)$:

\begin{equation}\label{eq_laplacian}
L(i, j)=\left\{\begin{array}{cc}
{1,} & {{\rm if} \ {i=j,}} \\
{-\frac{1}{|\mathcal{N}_{i}|},} & {{\rm if } \ j \in \mathcal{N}_{i},} \\
{0,} & { {\rm otherwise.}}
\end{array}\right.
\end{equation}
we use $\alpha_{s} \sim \textit{U} [0,\alpha ] $ to control the level of Laplacian smoothing. For the coordinate matrix $\mathbf{V}_{wm} \in \mathbb{R}^{N\times 3} $ of watermarked vertices $\mathcal{V}_{wm} $, we calculate the the coordinate matrix $\mathbf{V}_{att}$ of attacked vertices $\mathcal{V}_{att}$  as :

\begin{equation}\label{eq_smoothing}
\mathbf{V}_{att}=\mathbf{V}_{wm} - \alpha_{s}  \mathbf{L} \mathbf{V}_{wm}.
\vspace{0em}
\end{equation}

\paragraph{Cropping.} We first normalize the vertices in a unit square and search for the two farthest points in the negative quadrant and the positive quadrant respectively. Then We connect two points and simulate using a knife cutting perpendicular to the line. So that we can cut off the part of the mesh, with $\beta$ to control the minimum ratio of the reservation. $\beta_{c}\sim\textit{U}[\beta,1] $ is used to denote the actual ratio of the reservation at each cropping operation. 

We set the hyperparameters as follows: $\theta=15^{\circ}, \sigma=0.03, \alpha=0.2, \beta=0.8 $. Besides four attacks, we also integrate one identity layer which doesn't have any attack, to ensure the performance when no attack is suffered. During training, we randomly select one attack as the attack layer in each mini-batch. Then we can generate the attacked mesh $\mathcal{M}_{att}=(\mathcal{V}_{att},\mathcal{F}_{att})$ after the watermarked mesh $\mathcal{M}_{wm}=(\mathcal{V}_{wm},\mathcal{F}_{wm})$ passes through the attack layer.

\begin{figure*}[hbt]
\begin{center}
{\centering\includegraphics[width=1\linewidth]{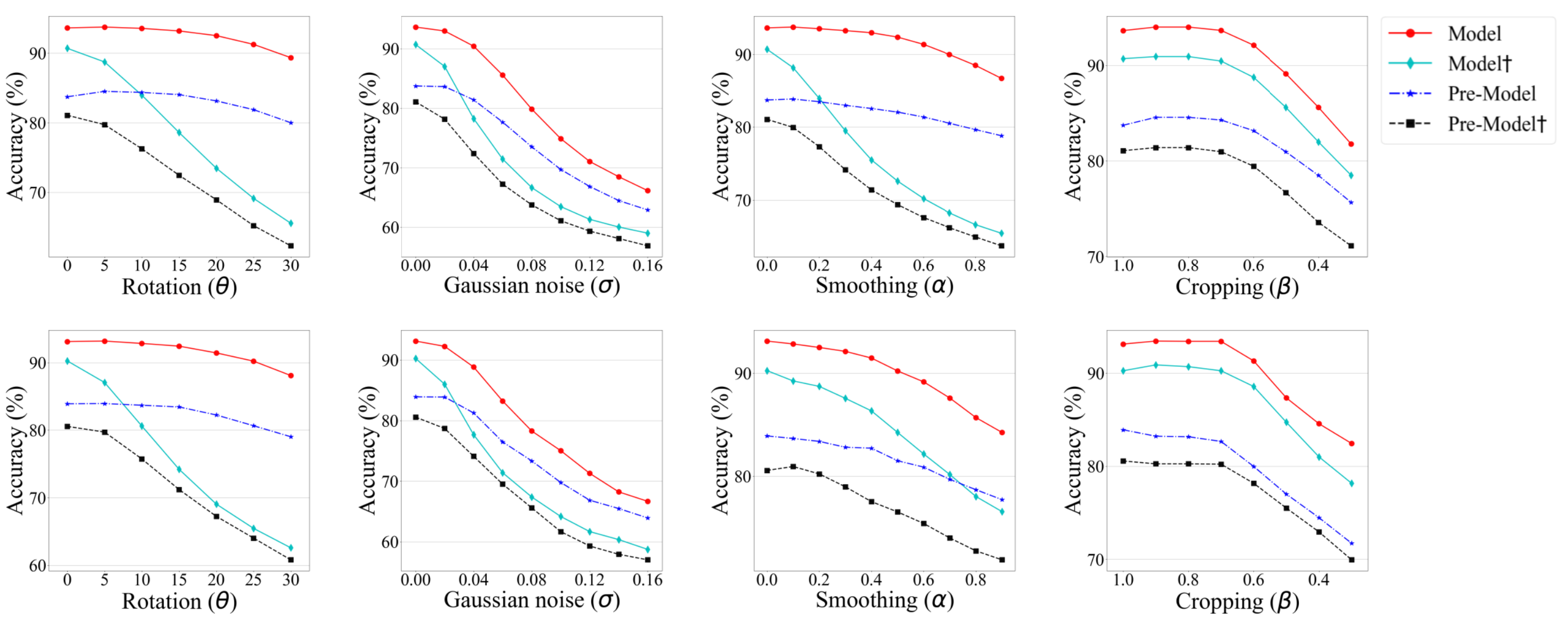}
}\\
\caption{Bit accuracy rate (\%) under different attacks and intensities on remeshed Asiadragon (top) and remeshed Hand (bottom) dataset. Model (red), Model$\dag$ (cyan), Pre-Model (blue) and Pre-Model$\dag$ (black) are showed. $\dag$ denotes training without the attack layers, Model represents the model trained on the remeshed dataset, and Pre-Model represents the model trained on the original dataset.}
\label{acc_remesh}
\end{center}
\vspace{-0em}
\end{figure*}

\begin{figure}[htb]
\begin{center}
{\centering\includegraphics[width=1\linewidth]{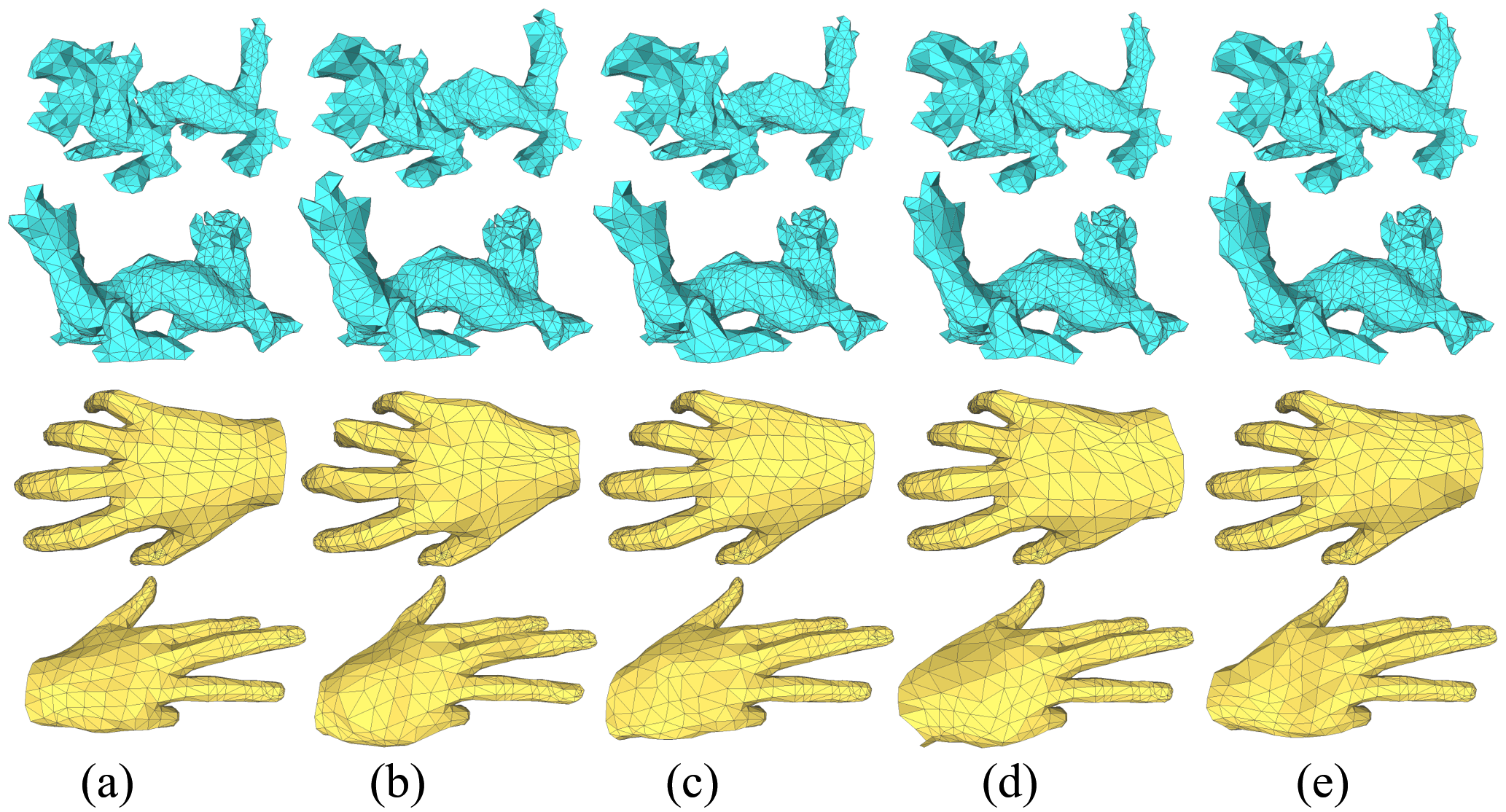}
}\\
\caption{Visual quality on remeshed Asiadragon (top two rows) and remeshed Hand (bottom two rows) dataset. (a) Original; (b) Model; (c) Model$\dag$; (d) Pre-Model; (e) Pre-Model$\dag$. $\dag$ denotes training without the attack layers, Model represents the model trained on the remeshed dataset, and Pre-Model represents the model trained on the original dataset.}
\label{qualitative-comparison-remesh}
\end{center}
\vspace{-2em}
\end{figure}

\section{Performances on Non-Template-Based Datasets and the Transferability Discussion}

We independently remesh each shape of Hand and Asiadragon dataset to $1024$ vertices by Trimesh library. So that each dataset is made up of non-template-based meshes (they don't share a fixed topology). Then we train our network using each remeshed dataset. Meanwhile, using new dataset, we also retrain the network without the attack layers (labelled with $\dag$). Besides, to test the transferability of our method, we also test our pre-trained model on the remeshed dataset, which is trained with original Hand dataset and Asiadragon dataset. For the sake of distinction, we use Model to denote the model trained on the remeshed dataset, and use Pre-Model to denote the model trained on the original dataset.


Table \ref{quantitative_comparison_remesh} shows the quantitative results. As our GCN is topology-agnostic, when trained with the remeshed dataset, we can still get the similar performance on imperceptibility and robustness. We can get $93.25\%$ of accuracy rate on remeshed Asiadragon dataset and $91.78\%$ of accuracy rate on remeshed Hand dataset. And the visual quality can also be guaranteed as shown in Figure \ref{qualitative-comparison-remesh}. For Pre-Model, we can still guarantee comparable performance when tested on remeshed datasets. There is still $83.63\%$ of accuracy rate on remeshed Asiadragon dataset and $83.00\%$ of accuracy rate on remeshed Hand dataset. And in terms of MRMS, HD and $l_{cur}$, the performance of Pre-Model is as similar as Model. In addition, we can find that the attack layers can help improve the accuracy both in Pre-Model and Model.

As shown in Figure \ref{acc_remesh}, we test the bit accuracy under each attack with different intensities on remeshed datasets. Under each attack, Pre-Model exhibits the similar curve property to Model, with about $10\%$ of accuracy rate reduction. Pre-Model can keep more than $78\%$ accuracy rate under rotation attack, and more than $75\%$ accuracy rate under smoothing attack.

\section{The Importance of Normalization}

\begin{figure}[t]
\begin{center}
{\centering\includegraphics[width=1\linewidth]{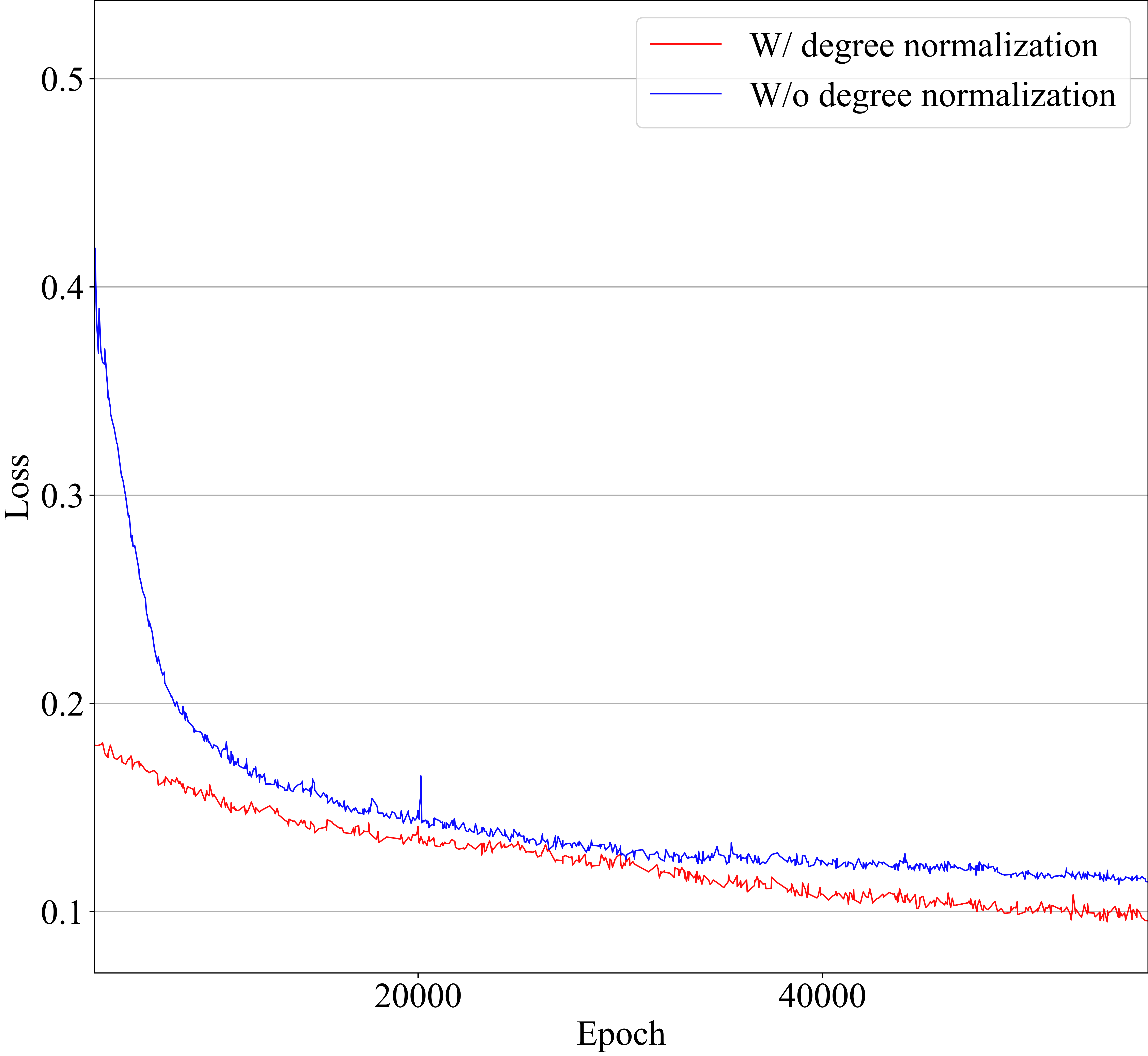}
}\\
\caption{The loss curve for the network trained on Hand dataset. The horizontal axis represents epoch number during training, and the vertical axis represents the total loss.}
\label{degree_normalization}
\end{center}
\end{figure}

For the design of topology-agnostic GCN, we employ degree normalization and batch normalization to help the convergence. The importance of both normalization operations has been demonstrated useful. Empirically, without the batch normalization, the mean values of the feature map rapidly swell and lead to training failure. The following equation represents the GraphConv operation, and $\frac{1}{|\mathcal{N}(i)|}$ denotes the degree normalization. When we don't use the degree normalization, the total loss drops a little more slowly as shown in Figure \ref{degree_normalization}.

\begin{equation}\label{eq_normalization}
f_{i}^{l+1}=w_{0}f_{i}^{l}+w_{1} \sum\limits_{j\in\mathcal{N}(i)} \frac{1}{|\mathcal{N}(i)|} f_{j}^{l},
\vspace{0em}
\end{equation}

\bibliography{reference}